\documentclass[12pt,preprint]{aastex}
\usepackage{subfigure}
\usepackage{amsmath}
\begin{document}

\title{Energy spectra of cosmic-ray nuclei  at high energies}


\author{H.\,S. Ahn\altaffilmark{1}, 
P. Allison\altaffilmark{2}, %
M.\,G. Bagliesi\altaffilmark{3},% 
L. Barbier\altaffilmark{4}, %
J.\,J. Beatty\altaffilmark{2},% 
G. Bigongiari\altaffilmark{3},% 
T.\,J. Brandt\altaffilmark{2},% 
J.\,T. Childers\altaffilmark{5}, 
N.\,B. Conklin\altaffilmark{6},
S. Coutu\altaffilmark{6}, 
M.\,A. DuVernois\altaffilmark{5},
O. Ganel\altaffilmark{1}, 
J.\,H. Han\altaffilmark{1}, 
J.\,A. Jeon\altaffilmark{7}, 
K.\,C. Kim\altaffilmark{1}
M.\,H. Lee\altaffilmark{1}, 
P. Maestro\altaffilmark{3,*}, 
A. Malinine\altaffilmark{1}, 
P.\,S. Marrocchesi\altaffilmark{3}, 
S. Minnick\altaffilmark{8}, 
S.\,I. Mognet\altaffilmark{6}, 
S.\,W. Nam\altaffilmark{7}, 
S. Nutter\altaffilmark{9},
I.\,H. Park\altaffilmark{7}, 
N.\,H. Park\altaffilmark{7}, 
E.\,S. Seo\altaffilmark{1,10}, 
R. Sina\altaffilmark{1}, 
P. Walpole\altaffilmark{1}, 
J. Wu\altaffilmark{1}, 
J. Yang\altaffilmark{7}, 
Y.\,S. Yoon\altaffilmark{1,10}, 
R. Zei\altaffilmark{3},  
S.\,Y. Zinn\altaffilmark{1}
}
\altaffiltext{1}{\footnotesize{Institute for Physical Science and Technology, University of Maryland, College Park, MD 20742, USA}}
\altaffiltext{2}{\footnotesize{Department of Physics, Ohio State University, Columbus, OH 43210, USA }}
\altaffiltext{3}{\footnotesize{Department of Physics, University of Siena and INFN, Via Roma 56, 53100 Siena, Italy }}
\altaffiltext{4}{\footnotesize{Astroparticle Physics Laboratory, NASA Goddard Space Flight Center, Greenbelt, MD 20771, USA}}
\altaffiltext{5}{\footnotesize{School of Physics and Astronomy, University of Minnesota, Minneapolis, MN 55455, USA}}
\altaffiltext{6}{\footnotesize{Department of Physics, Penn State University, University Park, PA 16802, USA }}
\altaffiltext{7}{\footnotesize{Department of Physics, Ewha Womans University, Seoul 120-750, Republic of Korea }}
\altaffiltext{8}{\footnotesize{Department of Physics, Kent State University, Tuscarawas, New Philadelphia, OH 44663, USA}}
\altaffiltext{9}{\footnotesize{Department of Physics and Geology, Northern Kentucky University, Highland Heights, KY 41099, USA}} 
\altaffiltext{10}{\footnotesize{Department of Physics, University of Maryland, College Park, MD 20742, USA}}
\altaffiltext{*}{Corresponding author. {\em E-mail address:} paolo.maestro@pi.infn.it (P. Maestro)}


\begin{abstract}
We present new measurements of the energy spectra of cosmic-ray (CR) nuclei
from the second flight of  the balloon-borne experiment 
Cosmic Ray Energetics And Mass (CREAM). 
The instrument  included different particle detectors 
to provide  redundant charge identification and  measure
the energy of CRs up to several hundred TeV. 
The measured individual energy spectra of C, O, Ne, Mg, Si, and Fe
are presented up to $\sim 10^{14}$ eV. The spectral shape 
looks nearly the same for
 these primary elements and it can be fitted to an $E^{-2.66 \pm 0.04}$
 power law in energy.
Moreover, a new measurement of the absolute intensity of nitrogen 
in the 100-800 GeV/$n$ energy range
with smaller errors than previous observations, 
clearly indicates a hardening of the  spectrum at high energy.  
The relative abundance of N/O 
at the top of the atmosphere  
is measured to be $0.080 \pm 0.025\, $(stat.)$\, \pm 0.025\,$(sys.)
at $\sim\,$800 GeV/$n$, in good agreement with 
a recent result from the first CREAM flight. 
%and a previous analysis with the CREAM-II data.
\end{abstract}

\section{Introduction}
Experimental studies of charged cosmic rays (CRs)
are focused on the  understanding of the  acceleration mechanism of high-energy 
CRs, identification of their sources, and 
 clarification of their interactions with the interstellar medium (ISM).
The origin of CRs is still under debate, although
 direct measurements  with instruments on stratospheric balloons or in space
have provided  information on their elemental composition and 
energy spectra,  
and indirect detection by ground-based experiments has extended  the measurements
up to the end of the observed all-particle spectrum. \\
It is generally accepted that CRs are
accelerated in blast waves of supernova remnants (SNRs), 
which are the only galactic candidates known with sufficient energy output 
to sustain the CR flux \citep{Bell, hillas}. 
Recent observation 
of emission of TeV gamma-rays from SNR RX J1713.7-3946 by the HESS Cherenkov telescope array
proved that high-energy charged particles 
are accelerated in SNR shocks  up to energies 
beyond 100 TeV \citep{HESS}. This result is consistent
with  expectations of the class of theoretical models that predict 
the existence of a rigidity-dependent limit, above which the diffusive 
shock acceleration becomes inefficient. 
The maximum energy attainable by a nucleus of charge $Z$ 
may range from $Z\times$10$^{14}$ eV  
to $\sim$ $Z\times$10$^{17}$ eV
depending on the model and types of supernovae considered \citep{LC, Berezhko, parizot}.
In this scenario, 
the energy spectra of elements exhibit a $Z$-dependent cutoff. As
 a consequence, the CR elemental composition is expected to change as a function of energy, marked by a  depletion 
of low-$Z$ nuclei in the several hundred TeV region. 
This mechanism has been proposed as a possible explanation of the steepening (``knee'')
in the CR energy spectrum (described by a power law: $dN$/$dE \propto E^{-\gamma}$)
with a change of the spectral index 
from $\gamma\approx\,$2.7 to $\gamma\approx\,$3.1  observed at energies around
4 PeV.
An alternative approach is adopted by models that relate the knee to leakage of CRs from the Galaxy.
In this case, the knee is expected to occur at lower energies for light elements  as compared to heavy nuclei, due to the 
rigidity-dependence  of the Larmor radius
of  CR particles  propagating in the galactic magnetic field  \citep{horandel2}.\\
A detailed understanding of the propagation of CRs during their wanderings inside the 
Galaxy and their interactions 
with the ISM is needed to infer  the injection spectra  of 
individual elements at the source from the observed spectra at Earth.
CR nuclei can be divided into primaries (i.e., H, He, C, O, Fe, etc.) that come 
from CR sources, and secondaries (such as Li, Be, B, Sc, Ti, V, etc.), which are  products 
of the interactions of primary nuclei with the ISM. 
The amount of material traversed by CRs between injection and observation can 
be derived from the measured ratio of secondary-to-primary  nuclei, 
such as the boron-to-carbon ratio (B/C) or the ratio of sub-iron elements to iron.
Likewise, their confinement time 
in the Galaxy can be determined through measurements of long-lived radioactive secondary nuclei \citep{longair, yanasak}. 
Observations from space-based experiments like CRN 
on the Spacelab2 mission of the Space Shuttle \citep{CRN2} and  C2 \citep{HEAO}
and Heavy Nuclei Experiment (HNE) \citep{binns} onboard the {\em HEAO 3} satellite
have shown that the energy spectra of 
the secondary nuclei  are steeper than those of the  primaries.
The escape pathlength of CRs from the Galaxy 
depends on the  magnetic rigidity as
 $R^{-\delta}$, with the parameter $\delta\approx\,$0.6. 
This implies that the spectra observed at Earth are steeper  than the injection spectra, 
i.e., the spectral index at the source is smaller by the value $\delta$. \\
Those pioneering measurements were statistically limited at energies of order $\sim$ 10$^{11}$ eV.
Accurate direct measurements of the energy spectra of individual elements into the knee region 
are needed to discriminate 
between different astrophysical models proposed to explain the acceleration and propagation mechanisms.
Recently, a new generation of balloon-borne experiments
has performed accurate measurements of CRs in the previously experimentally unexplored 
TeV region \citep{ATIC, TRACER}. Among these, 
the Cosmic Ray Energetics And Mass (CREAM) 
experiment was designed to  measure directly
the elemental composition  and the energy spectra of 
CRs from hydrogen to iron
over the energy range 10$^{11}$-10$^{15}$ eV in a series of flights \citep{seo}. 
Since 2004, four instruments were  successfully flown on long-duration balloons
in Antarctica. The instrument configurations varied slightly in each mission
due to various detector upgrades.
The first flight produced important results 
extending the measurement of the relative abundances of CR secondary nuclei B and N close to the  
highest  energies ($\sim$ 1.5 TeV/$n$) allowed by the irreducible background
generated by the residual atmospheric overburden at balloon altitudes.
The data clearly indicate that the escape pathlength of CRs from the Galaxy 
has a power-law rigidity-dependence, with the parameter $\delta$ in the range 0.5-0.6 \citep{Ahn2008}.  

In this work, we present new  measurements of the energy spectra of the major primary CR nuclei
from carbon to iron, up to a few TeV/$n$ made with the  second CREAM flight (CREAM-II). A
new measurement of the nitrogen intensity with unprecedented statistics
is also presented up to 800 GeV/$n$. 
The measured N/O ratio confirms the results of CREAM-I. 
%and  a previous analysis of CREAM-II data \citep{AhnSubmit}.
The procedure used to analyze the data 
and reconstruct the CR energy spectra is described in detail
together with an assessment of the systematic uncertainties.\\
%%%%%%%%%%%%%%%
\section{The CREAM-II instrument}
\begin{figure}
\begin{center}
\includegraphics[angle=90,scale=.65]{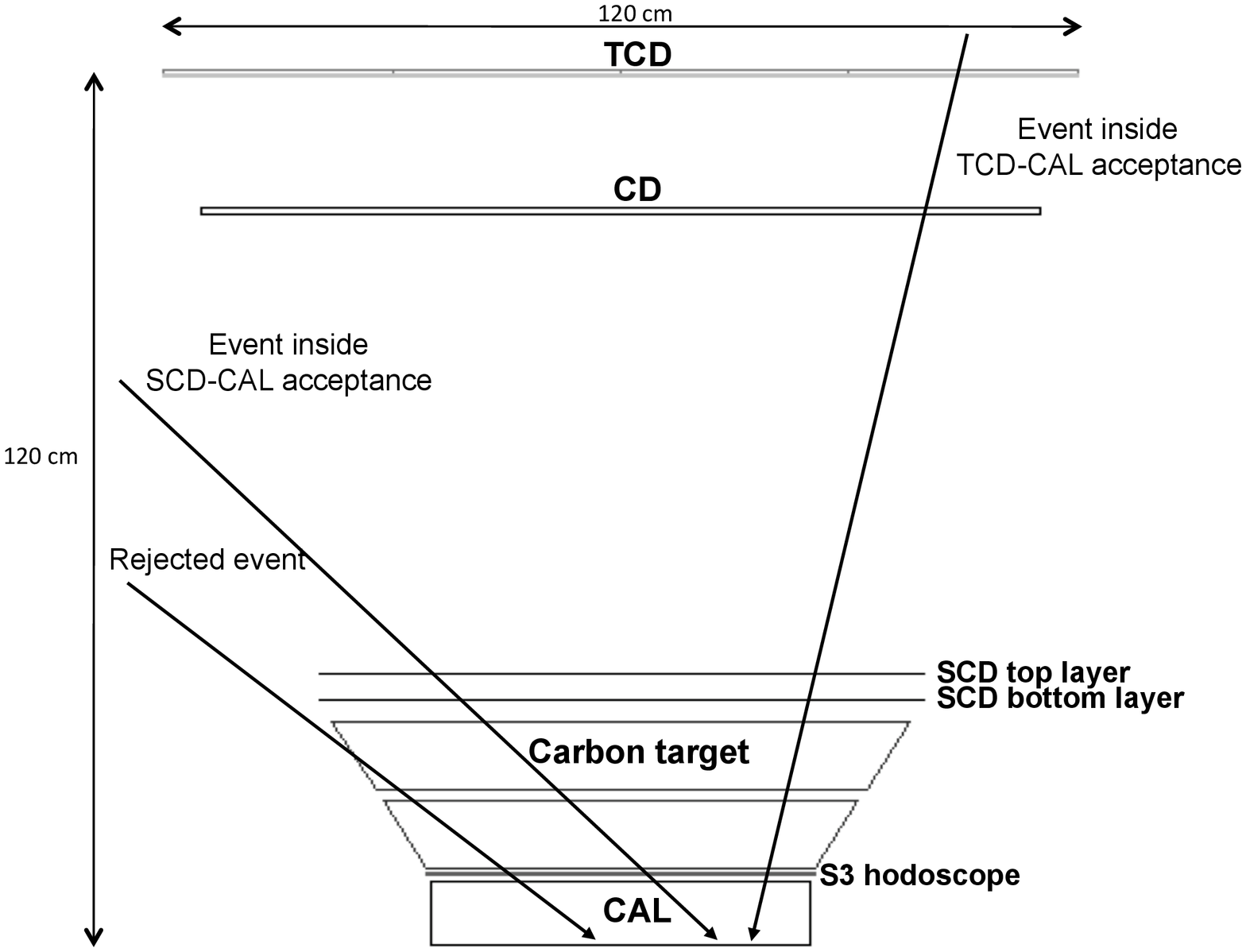}
\caption{Schematic view of the CREAM-II instrument.}
\label{CREAM2view}
\end{center}
\end{figure}

The instrument for the second flight  included: a
redundant system for particle identification, consisting 
(from top to bottom)
of a timing-charge detector (TCD), a
Cherenkov detector (CD), a pixelated silicon charge detector (SCD), and  
a sampling imaging calorimeter (CAL) designed to provide a
measurement of the energy of CR nuclei in the multi-TeV region.
Figure \ref{CREAM2view} depicts the detector arrangement.\\
The TCD is made of two orthogonal layers of four 5 mm-thick plastic scintillator paddles each, 
covering an area of 120$\times$120 cm$^2$. The paddles are read out by fast 
photomultiplier tubes (PMTs) via twisted-strip adiabatic light guides.
A custom design of the electronics combines fast peak detection with a threshold timing measurement of the 
PMTs signals in order to determine each element's charge  
with a resolution $\lesssim$ 0.35 $e$  (in units of the electron charge) and to
discriminate against albedo particles.
A hodoscope (S3) of 2$\times$2 mm$^2$ square scintillating fibers read out by two PMTs
is located just above the calorimeter, and
provides a reference time for TCD timing readout. Further details about the TCD design
and performance can be found in \citet{Ahn2009}.
The CD is a 1 cm-thick plastic radiator with 1 m$^2$ surface area, instrumented with eight PMTs 
viewing wavelength shifting bars
placed along the radiator edges. 
It is used to flag relativistic particles
and to provide a charge determination complementary to that of the TCD.\\
The SCD is comprised of two layers of 156 silicon sensors each.
Each 380 $\mu$m-thick sensor is segmented into an array of 4$\times$4 
pixels, each of which has an active area of 2.12 cm$^2$ \citep{Park}. 
The sensors are slightly tilted and overlap each other in both lateral directions, 
providing a full coverage in a single layer, with a 77.9 $\times$ 79.5 cm$^2$ area.
The dual layers, about 4 cm apart,
cover an effective inner area of 0.52 m$^2$ with no insensitive regions.
The overall SCD vertical  dimension, including the ladders, the mechanical
support structure and the electromagnetic shielding cover, 
is 97.5 mm  \citep{Nam}.
The readout electronics of the 4992 pixels is based on a 16-channel ASIC chip  
followed by 16-bit analog-to-digital converters (ADCs), providing a fine charge resolution over a wide dynamic range 
from hydrogen to nickel.\\
%The dual layers, about 4 cm apart,
%cover an effective area of 0.52 m$^2$ with no insensitive regions.
The CAL is a stack of 20 tungsten plates, each 1 radiation length ($X_0$) thick
and with surface area 50$\times$50 cm$^{2}$, 
interleaved with active layers  of 1 cm-wide and 50 cm-long scintillating-fiber
ribbons. Each ribbon is built by gluing together 19  scintillating fibers of 0.5 mm diameter each.
The light signal from each ribbon is collected  by means of an acrylic light mixer
coupled to a bundle of 48 clear fibers.
This is split into three sub-bundles (with 42, 5 and 1 fibers, respectively), each
feeding a pixel of a hybrid photodiode (HPD).  
This optical division is used to match the wide dynamic range of the calorimeter 
to that of the front-end electronics, providing three readout scales (low, medium, and high) 
with different sensitivity.
A total of 2560 channels are readout from 40 HPDs powered in groups of 5 units.\\
%In this way the dynamic
%range of the calorimter is divided into 3 sub-ranges 
%The light signals from the ribbons are collected by means of acrylic light mixers 
%coupled to clear fiber bundles, and are read out by 40 hybrid photodiodes (HPDs). 
The longitudinal and lateral segmentations
of the CAL correspond to 1 $X_0$ and 1 Moliere radius, respectively.
The ribbon planes are alternately oriented along orthogonal directions
to image in three dimensions, the development of  showers produced by the incoming nuclei which 
interact inelastically in the $\sim$ 0.46 ${\lambda}_{int}$-thick carbon target preceding the CAL  \citep{Marrocchesi2}.
The finely grained sampling calorimeter can track, by reconstructing the shower axis, the
incident particle trajectory with enough accuracy 
to determine which segment of each charge detector it traversed,
and thereby allowing discrimination against segments hit by the backscattered particles produced in the shower. 
This feature is essential for
a reliable and unambiguous charge identification.
%%%%%%%%%%%%%%%%%%%%%%%%%%%%%%%%%%%%%%%%%%%%%%
\section{Balloon flight and instrument performance}
The second CREAM instrument was launched from McMurdo station 
in Antarctica on 2005 December 16th. The balloon circled the South Pole twice at a
float altitude between 35 and 40 km.
The flight was terminated on 2006 January 13th, 28 days after the launch. 
During the whole flight, 
the instrument housekeeping system monitored online
temperatures, currents, voltages of the front-end electronics boards, and the high-voltage supply values of 
the HPDs and PMTs.
The performance of all sub-detectors
met the design technical specifications.
The CAL and TCD high-voltage systems  worked successfully at low atmospheric
pressure.
The thermal behavior of the various detectors was found to be in 
good agreement with  expectations from  thermal models. 
The temperature of the instrument crates stayed within the required operational range
with daily variation of a few $^\circ$C, depending on the inclination of the Sun
during each 24 hr cycle.
Front-end electronics channels showed stable gains over the whole flight.
Pedestal values of the 2560 CAL and 4992 SCD channels
were collected automatically every 5 minutes
in order to monitor their drift  as a function of the detector temperature
and to perform accurate pedestal subtraction. 
1.7\% of the SCD  and 0.7\% of the CAL channels were 
noisy or inefficient and were masked off.\\
During the flight, 
the data acquisition was enabled
whenever a shower developing in at least six planes was detected in the CAL (CAL trigger) and/or
a relativistic CR with $Z\geq$ 2 was identified by the TCD and CD  (TCD trigger). 
The high-energy data, i.e., the CAL triggered events, were transmitted via Tracking and Data Relay
Satellite System (TDRSS), while the low-energy data, i.e., events triggered only by the 
TCD and not by the CAL, were recorded on an onboard disk. 
A total of 57 GB of data were collected.
In order not to saturate the disk space too early in the flight, 
the highly abundant low-energy particles triggered only by the TCD
were prescaled, so that only a fraction of these events were recorded. 
This fraction of recorded
TCD triggers (prescale factor) could be set 
by the data acquisition program and changed during the flight depending on the trigger rate.
On average, only one every six TCD triggered events was retained.
A detailed description of the instrument  performance during the second 
flight may be found in \citet{Marrocchesi}.
\section{Data analysis}
The present analysis uses a subset of data collected 
from December 19th to January 12th, 
under stable instrument conditions,
representing a total of
24 days' worth of data taking (e.g., removing periods early
in the flight and at the very end, when the instrument was either 
being adjusted and optimized, or prepared for flight termination).
The first step in the analysis is to correct 
the  raw data for 
the gain variations among the channels and 
the drift of the pedestals with temperature.
Then the trajectory of each CR through the instrument
is reconstructed and each event  assigned a charge and an energy. 
At this point, the events of each charge species are sorted into energy intervals
and the number of counts in each bin is corrected for several effects
in order to compute the differential intensity.
The various steps of the analysis procedure are described 
in detail in the following subsections.
\subsection{Trajectory reconstruction}
To  determine
the arrival direction of a CR, 
the axis of the shower imaged by the calorimeter is measured  with the following method.
In each CAL plane, an offline software algorithm searches for clusters of adjacent cells 
with pulse heights larger than a threshold value,
corresponding to an energy deposit of about 10 MeV.
The  cluster formed by the cell 
with the maximum pulse height and its two neighbors
defines a candidate track point. Its coordinates are computed as 
the center of gravity of the cluster. 
A track is formed by matching 
at least three candidate track points 
sampled in each CAL view. 
The shower axis parameters are calculated by a linear $\chi^2$ fit of the track.
Track quality is assured by requiring a value of  $\chi^2<\,$10.
The reconstructed shower axis is projected back to the top of the instrument
in order to determine 
which SCD pixels and TCD paddles were hit by the incoming CR. 
Particles showering in the CAL, but crossing neither the 
TCD scintillator paddles nor the SCD planes are rejected.  
Otherwise, if the particle crosses the SCD, 
a circle of confusion with a 3 cm radius centered 
on the impact point of the track is traced in each SCD layer.
The value of the radius is equal to about three times 
the estimated uncertainty on the impact point.
The hit pixels within each circle of confusion are scanned 
and the pixel with the maximum pulse height is selected for the charge assignment procedure.
In order to improve the trajectory determination and the accuracy of the pathlength correction, 
the two selected pixels,  one per SCD layer, are added to the candidate
 points and the track is fitted again.
In this way, the intersection 
of the particle trajectory with the top SCD layer is 
measured with a spatial resolution of 7 mm rms. 
The incidence direction is determined with an accuracy of better than  1$^{\circ}$
as estimated with Monte Carlo (MC) simulations. \\
This procedure reconstructs well the trajectories of over 95\% of the CR particles
with energy $>$ 3 TeV  
triggered by the CAL. 
It can also  identify
the TCD triggered events
which produce 
a shower in the CAL, but with an energy release insufficient to satisfy the CAL trigger condition.
%%%%%%%%%%%%%%%%%%%%%
\subsection{Charge assignment}\label{chargepar}
\begin{figure}
\begin{center}
\subfigure[]
{
\includegraphics[scale=0.73]{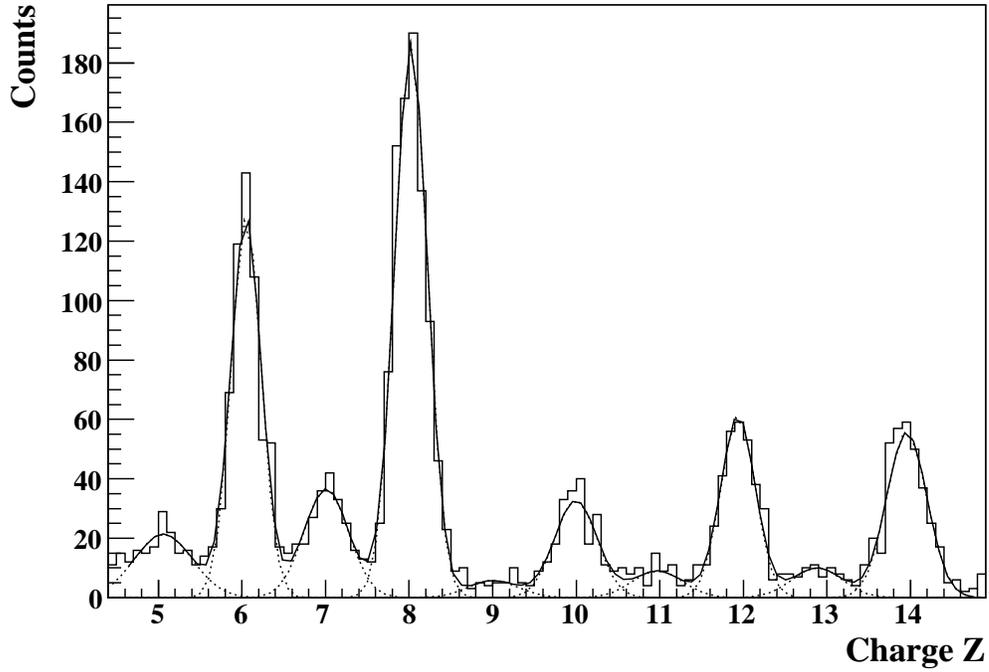}
\label{Zfiga}                            
}
\subfigure[]
{
\includegraphics[scale=0.73]{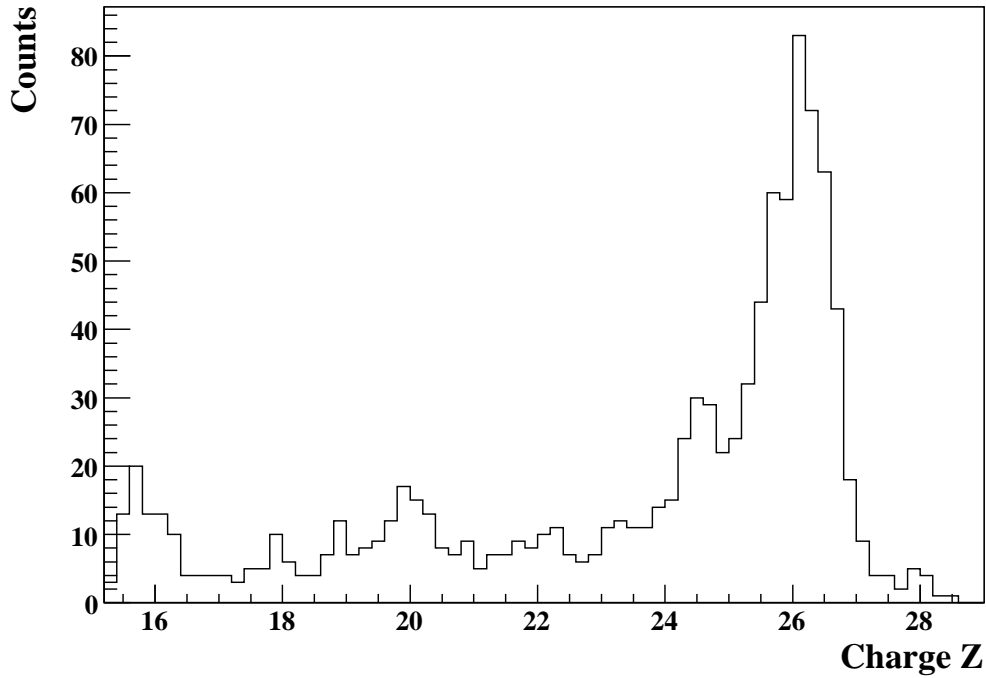}
\label{Zfigb}                            
}
\caption{Charge histogram obtained by the SCD in the elemental range (a) from boron to silicon and (b)  from sulfur to nickel. The distributions are only indicative of the SCD charge resolution and the relative elemental abundances are not meaningful. The peaks from B to Si are fitted to a multi-Gaussian function. }
\label{Zfig}
\end{center}
\end{figure}

The identification of the charge $Z$ of a CR relies
on two independent samples of its specific ionization $dE/dx$
provided by the SCD layers. 
Since $dE/dx$ depends on $Z^2$ and for solid absorbers
it is nearly constant in the relativistic regime, 
the particle charge $Z$ can be determined by measuring 
the amount of ionization charge a CR produces when traversing the 
silicon sensors. 
In order to avoid charge misidentification of the incident nucleus,
caused by the backscattered shower particles reaching the SCD,
and to obtain high-purity samples of CR elements, 
the signals, $S_{top}$ and $S_{bottom}$, of the two pixels selected by the tracking algorithm
are compared and required to be consistent within 20\%.
If this coherence cut is satisfied, 
they are corrected for the pathlength 
(estimated from the track parameters)
traversed by the particle in the silicon sensors, which is proportional to 
$1/\cos{\left(\theta\right)}$, $\theta$ being the angle between the particle trajectory and 
the axis of the instrument. 
A reconstructed charge $Z_{rec}$ is assigned to the particle by combining 
the ionization signals matched with the track on  both layers, according to the formula:
\begin{equation}
Z_{rec} = \sqrt{\frac{S_{top} + S_{bottom}}{2} \times \frac{\cos{\left(\theta\right)}}{s_0} }
\end{equation}
where $s_0$ is a calibration constant, inferred from the flight data,  
to convert the pulse heights into an absolute charge scale.
The reconstructed charge distribution is shown in Figure \ref{Zfig}.
The elemental range from boron to silicon is fitted with a multi-Gaussian 
function (Figure \ref{Zfiga}). From the fit, 
a charge resolution $\sigma$ is estimated  as:
0.2 $e$ for C, N, O and $\sim$ 0.23 $e$ for Ne, Mg, Si.
Though a multi-Gaussian fit in the elemental region from sulfur to nickel
cannot be done due the limited statistics,  
the charge resolution for iron is estimated to be
$\sim$ 0.5 $e$ from the width of the Fe peak.
A 2$\sigma$ cut around the mean charge value is applied to select samples 
of C, O, Ne, Mg and Si.
A cut of $Z_{rec} \pm 1 \sigma$
is imposed for Fe due to the corresponding lower SCD  resolution, 
while for N a cut of $Z_{rec} \pm 1.5 \sigma$ and 10\%
coherence  level between the selected pixels signals, 
are used to avoid contamination from more abundant adjacent charges.
%%%%%%%%%%%%%%%%%
\subsection{Energy measurement}
Due to the limitations on mass and size imposed to the payload, 
a balloon-borne experiment cannot employ  
a total containment hadronic calorimeter to measure the energy of CR nuclei.
An alternative and workable technique 
is to use a thin ionization calorimeter to sample the electromagnetic 
core of the hadronic cascade 
initiated by a CR interacting in a target preceding the calorimeter.
In fact, though
a significant part of the hadronic cascade energy  leaks out of the calorimeter,
the energy deposited in the calorimeter by the shower core still scales 
linearly with the incident particle energy, 
albeit with large event-to-event fluctuations.
As a result, the energy resolution is poor by the standards of hadron calorimetry
in experiments at accelerators. 
Nevertheless, it is sufficient to reconstruct the steep energy spectra of CR
nuclei with a nearly energy independent resolution.\\
The CREAM calorimeter was designed to measure CRs 
over a wide energy range, from tens of GeV up to about 1 PeV.
It is characterized by a very small sampling fraction; 
only about 0.1\% of the CR energy is converted into visible signals in the CAL active planes  
and collected by the photosensors.
Three different gain scales (low, medium and high)  were implemented for each ribbon in order to prevent
saturation of the readout electronics and thereby to ensure
a linear detector response up to very high energies.\\
The total energy 
deposited in the calorimeter by an interacting nucleus,
is measured by summing up the corrected pulse heights of all the cells. 
The cells are equalized for non-uniformity in 
light output and gain differences among the photodetectors
using a set of calibration constants 
from accelerator beam tests (described below), scaled according to the high-voltage settings during the flight.
Low and medium gain scales are inter-calibrated using the flight data; the medium-range signal of each ribbon 
is used whenever the corresponding low-range signal deviates from linearity due to saturation. 

The CREAM calorimeter was tested, equalized, and calibrated pre-flight at the CERN SPS accelerator
with beams of electrons and protons up to a few hundred GeV \citep{Ahn2004}.
Its response to relativistic nuclei was also studied by exposing the calorimeter to 
nuclear fragments from a  158 GeV/$n$ primary indium beam.
%Instead, the calorimeter of the first CREAM flight was also exposed to  
%a beam of ion fragments. 
The detector was found to have a linear response up
to the maximum  beam energy, equivalent to about 8.2 TeV particle energy,
and a nearly flat resolution (around 30\%), 
at energies above 1 TeV for all heavy nuclei with Z$\geq$5 \citep{Ahn}.
%Since the two calorimeters have a very similar design
%and comparable responses to electrons and protons, 
%the CREAM-I CAL response curve for heavy ions was applied to  the second flight data as well. 
Above the maximum beam energy, the calorimeter response was extrapolated 
using MC simulations, which predict a %nearly 
linear behavior up to hundreds of TeV.
%%%%%%%%%
\subsection{Energy deconvolution}
\begin{figure}
\epsscale{.90}
\plotone{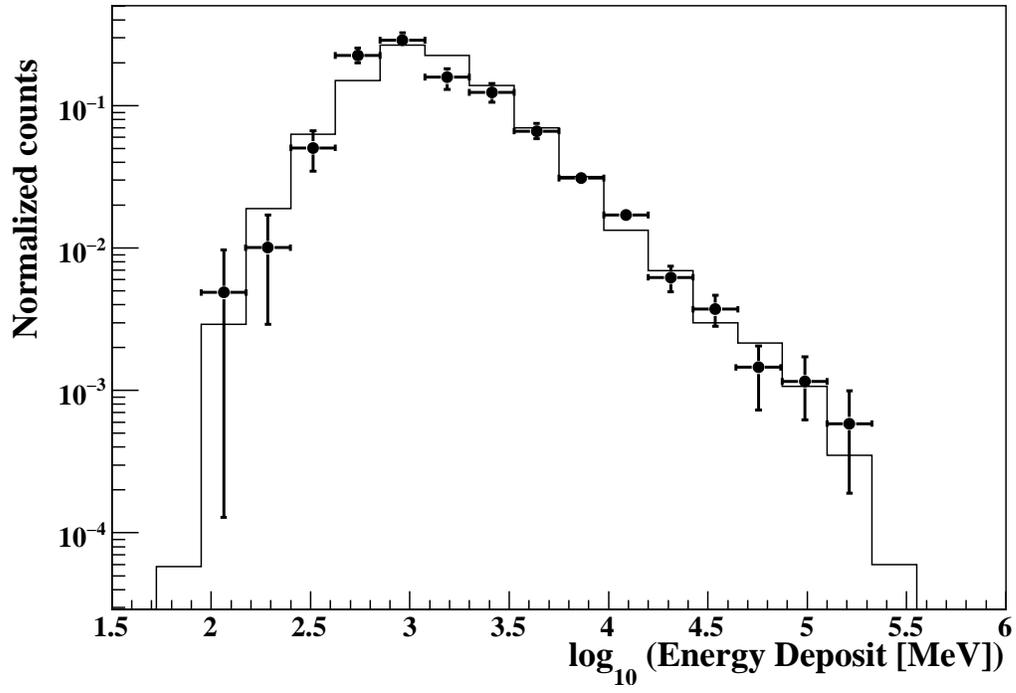}
\caption{Energy deposited in the calorimeter by a selected sample of carbon nuclei. Simulated (histogram) and real (dots)
events are shown.}
\label{CALedep}                            
\end{figure}

Once  each CR is assigned a charge and  energy, the reconstructed particles of each nuclear
species are sorted into energy intervals 
commensurate with the  rms resolution of the calorimeter. 
Due to the finite energy resolution of the detector, 
the measured number of events
in each energy bin must be corrected 
for overlap with the neighboring bins. 
This unfolding procedure
requires solving a set of linear equations
\begin{equation}
M_i = \sum_{j = 1}^{n} A_{ij} N_j \;\;\;\;\;\;\;\; i=1,..., m
\end{equation}
relating the 
``true'' counts $N_j$ in one of $n$ incident energy bins 
to the measured counts $M_i$ in one of $m$ deposited energy bins.
A generic element of the mixing matrix $A_{ij}$ represents 
the probability that a CR particle, carrying an energy 
corresponding to a given energy  bin \emph{j}, produces 
an energy deposit in the calorimeter falling into  bin \emph{i} instead. 
A detailed MC simulation of the instrument, 
based on the FLUKA 2006.3b package  
\citep{fluka}, was developed 
 to estimate the unfolding matrix. 
Sets of nuclei, generated isotropically and with energies chosen according to a power-law spectrum,
are analyzed with the same procedure
used for the flight data. Each matrix element $A_{ij}$
is calculated by correlating the generated spectrum with the distribution of the 
deposited energy in the calorimeter \citep{Zei}.
In order to get a reliable set of values of the unfolding  matrix for each nucleus, 
the MC simulation is finely tuned to reproduce both flight data and the calibration data
collected with accelerated particle beams.
The agreement of the MC description with the real instrument behavior was carefully checked.
As an example, in Figure \ref{CALedep}
the response of the calorimeter to carbon nuclei from the flight data is compared
with an equivalent set of simulated events.\\
The elements of the unfolding matrix depend on the spectral index assumed for
the generation of MC events. For each nucleus we used 
the spectral index value reported by \citet{horandel}, 
obtained by combining  direct and indirect CR observations.
In order to check the stability of our final results, 
the unfolding procedure was repeated
by scanning several values of the spectral index 
in the range $\pm\,$0.2 around the reference value.
No significant bias of the final results was observed.
%%%%%%%%%%%%%%%%%%%%%%%%%%%%%%%%%%%%%%%%%%%%%%%%%%%%%%%%%%%%%%%%%%%%%%%
\subsection{Corrections for interactions in the atmosphere and the detector}
\begin{figure}
\epsscale{.85}
\plotone{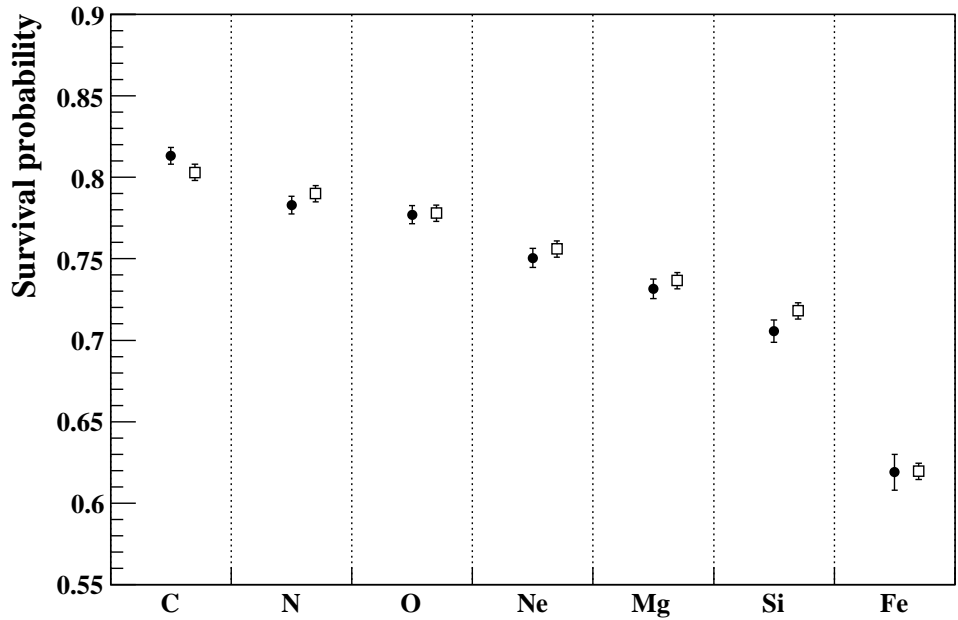}
\caption{Survival probabilities of CR nuclei traversing the materials between the TOI
 and the upper SCD layer. For each elemental species, the correction 
is estimated in two different ways: (filled circles)
from the FLUKA-based \citep{fluka} Monte Carlo simulation of the CREAM-II instrument;
(open squares) from calculations using the 
semiempirical total cross-section formulae 
for nucleus-nucleus reactions developed by \citet{sivher}.}
\label{TOI}
\end{figure}

\begin{figure}
\epsscale{.85}
\plotone{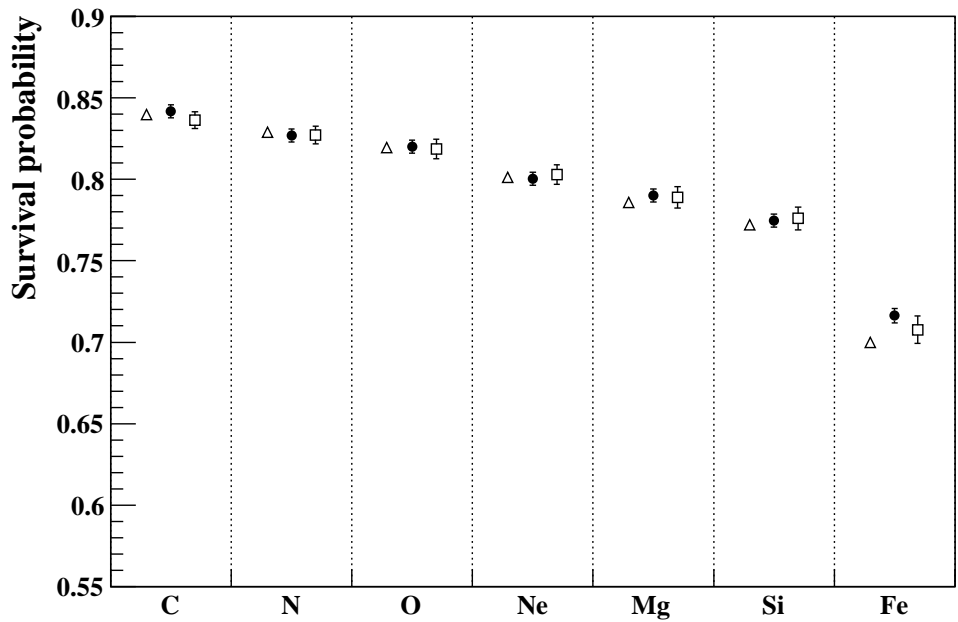}
\caption{Survival probabilities of nuclei in a residual atmospheric overburden of 3.9 g cm$^{-2}$.
For each elemental species, three values of the same correction
are shown, obtained respectively from 
the FLUKA-based \citep{fluka} Monte Carlo simulation of the CREAM-II instrument 
(filled circles),
and from  calculations using 
 two different  parameterizations of the total fragmentation 
cross sections for nucleus-nucleus reactions: \citet{sivher} (open squares); \citet{kox}  (open triangles).}
\label{TOA}
\end{figure}

CRs may undergo spallation reactions when traveling
through the atmosphere above the balloon or 
crossing the instrument material of CREAM above the SCD. 
This results in the 
generation of secondary nuclei with lower charge than the parent's, and of
 about the same energy per nucleon.
Then, for each nuclear species, inelastic interactions
lead to a loss of primary particles but also 
a gain of secondaries produced by heavier nuclei. 
For a correct determination of the CR fluxes, 
the data 
must be corrected for this effect.
A two-step procedure is used which
takes into account separately the CR
pathlength in air (correction 
to the top of the atmosphere (TOA))
and the amount of instrument material
encountered by the particle before reaching the SCD. The latter leads to a  correction to the top of the instrument (TOI).
For each step the corrections are calculated by solving the matrix equation 
\small
\begin{equation}
\begin{pmatrix} N_C \\ N_N \\ N_O \\ N_{Ne} \\ N_{Mg} \\ N_{Si} \\ N_{Fe}\end{pmatrix}_{\mathcal{F}} = 
\begin{pmatrix} S_C & F_{N \rightarrow C} & F_{O \rightarrow C} & F_{Ne \rightarrow C} & F_{Mg \rightarrow C} & F_{Si \rightarrow C} & F_{Fe \rightarrow C} \\ 0 & S_N & F_{O \rightarrow N} & F_{Ne \rightarrow N} & F_{Mg \rightarrow N}  & F_{Si \rightarrow N} & F_{Fe \rightarrow N} \\ 0 & 0 & S_O & F_{Ne \rightarrow O} & F_{Mg \rightarrow O} & F_{Si \rightarrow O}  & F_{Fe \rightarrow O} \\ 0 & 0 & 0 & S_{Ne} & F_{Mg \rightarrow Ne} & F_{Si \rightarrow Ne} & F_{Fe \rightarrow Ne} \\ 0 & 0 & 0 & 0 & S_{Mg} & F_{Si \rightarrow Mg} & F_{Fe \rightarrow Mg} \\ 0 & 0 & 0 & 0 & 0 & S_{Si} & F_{Fe \rightarrow Si} \\ 0 & 0 & 0 & 0 & 0 & 0 & S_{Fe} \end{pmatrix} \begin{pmatrix} N_C \\ N_N \\ N_O \\ N_{Ne} \\ N_{Mg} \\ N_{Si} \\ N_{Fe}\end{pmatrix}_{\mathcal{I}}
\end{equation} 
\normalsize
which describes the transport process of the nuclei 
from an incident level $\mathcal{I}$ (e.g., the top of the instrument)
to a final level $\mathcal{F}$ (e.g., the SCD).
The vector $\mathbf{N}_{\mathcal{I}}$ represents the number
of nuclei of each species counted in a given energy
interval, while the vector $\mathbf{N}_{\mathcal{F}}$ 
is the number of counts corrected for the charge-changing interactions
in the materials between the levels $\mathcal{I}$ and $\mathcal{F}$.
The diagonal elements of the transport matrix
represent the survival probabilities of each species $Z$ in the passage from $\mathcal{I}$ to $\mathcal{F}$, 
while a generic off-diagonal element $F_{Z_2 \rightarrow Z_1}$ gives the fraction of nuclei $Z_2$ which spallate 
to produce
nuclei of charge $Z_1$ ($< Z_2$).\\
The TOI transport matrix  is obtained from the MC simulation of the instrument.
The mass thickness of the CREAM-II materials
above the SCD is estimated to be $\sim\ $4.8 g cm$^{-2}$.
The fraction of surviving nuclei, i.e., the diagonal matrix elements, 
ranges from 81.3\% for C to 61.9\% for Fe.
These MC values
are compared with the ones calculated using the 
semiempirical total cross-section formulae 
for nucleus-nucleus reactions given in \citet{sivher}. 
Good agreement, within 1\%, is found as shown in Figure \ref{TOI}.\\
The TOA matrix elements are calculated  by simulating with FLUKA  the atmospheric
overburden during the flight (3.9 g cm$^{-2}$ on average). 
Since the probability of interaction depends on 
the nucleus pathlength in air, the simulated events are generated 
according to the measured
zenith angle distribution of the reconstructed nuclei, which is peaked 
around 20$^{\circ}$.
Survival probabilities ranging from 84.2\% for C to
71.6\%  for Fe  are found (Figure \ref{TOA}). They turn out to be in good
agreement with the values calculated from two different empirical parameterizations \citep{sivher,kox}
of the total fragmentation cross sections of heavy relativistic nuclei in targets of different mass.\\% of nuclei on a 3.9 g/cm$^2$-thick slab of air.
The off-diagonal elements $F_{Z_2 \rightarrow Z_1}$ of both the TOI and TOA transport matrices are of order 1\%-2\%. The comparison between values obtained from different models is discussed in Section \ref{syspar}. \\
%%%%%%%%%%%%%%%%%%%%%%%%%%%%%%%%%%%%%%%%%%%%%%%%%%%%%%%%%%%%%%%%%%%%%%%
\subsection{Absolute differential intensity calculation}
The absolute differential intensity  $dN/dE$ at an energy $\hat{E}$
is calculated for each elemental species  
% by dividing 
%the number of events $\tilde{N_i}$, obtained from the unfolding algorithm
%and corrected to the top of the atmosphere, 
% by the energy bin width $\Delta E_i$, 
according to  the formula
\begin{equation}
\frac{dN}{dE}(\hat{E}_i) = \frac{\tilde{N_i}}{\Delta E_i}\times \frac{1}{\epsilon \times S\Omega \times T }
\label{eq_flux}
\end{equation}
where $\tilde{N_i}$ is the number of events  obtained from the unfolding algorithm
and corrected to the TOA, $\Delta E_i$ is the energy bin width, 
$T$ is the exposure time, $S\Omega$ the geometric factor 
of the instrument, and $\epsilon$ the efficiency  of the analysis selection. 
Since a significant fraction 
of the reconstructed events at energies $<$3 TeV 
did not satisfy the CAL trigger condition and 
were triggered only by the TCD,  
the number of these events in the first energy intervals
must be further divided by the prescale factor (1/6 on average during the flight)
of the TCD-based trigger system. 

{\em Median energy}. According to the work of \citet{LW}, 
each data point, i.e., each differential intensity value, 
is centered at a median energy $\hat{E}$, defined as the energy at which the measured
spectrum is equal to the expectation value of the ``true'' spectrum. 
For a power-law spectrum ($E^{-\gamma}$) with spectral index $\gamma$, the median energy
is calculated as
\begin{equation}
\hat{E} = \left[ \frac{E_{2}^{1 - \gamma} - E_{1}^{1 - \gamma}}{\left(E_{2} - E_{1} \right) \left(1 - \gamma \right)} \right]^{-\frac{1}{\gamma}}
\end{equation}
where $E_{1}$ and $E_{2}$ are the lower and upper limits of a given energy bin.
If  the highest energy interval is not  limited, i.e., $E_2 \rightarrow \infty$,
the median energy is $\hat{E} =  2^{\frac{1}{\gamma - 1}}\, E_1$.
In this case, the energy interval is an integral bin and its width (to be used in Equation \ref{eq_flux}) is
calculated as
\begin{equation}
\Delta E = \frac{E_1}{2^{\frac{\gamma}{1 - \gamma}}\, (\gamma - 1)}
\end{equation}
Since $\gamma$ is not known a priori, an iterative procedure is performed 
to compute the median energy of each bin. For each element,  
the spectral index reported by
\citet{horandel} is used as the initial  value of $\gamma$.
We found that  $\hat{E}$
depends weakly on $\gamma$ for a variation of $\pm\,$0.1 around the 
initial value. 

{\em Geometric factor}. The geometric factor $S\Omega$ is estimated from MC simulations
by counting the fraction of generated particles 
entering the trigger-sensitive part of
instrument. Two groups of events are distinguished as shown in Figure \ref{CREAM2view}: CRs 
crossing both the TCD and the SCD before impinging onto the CAL (named ``golden'' events), and particles
entering the instrument through the side 
(i.e., crossing neither the TCD nor the CD) 
and traversing only the SCD and the CAL.
The estimated acceptance is (0.19 $\pm$ 0.01) m$^2$sr in the first case, while 
including both  event types yields a value of (0.46 $\pm$ 0.01) m$^2$sr.
Both  values are charge and energy independent.
For the second group of events, the corrections to the TOI are not necessary.

{\em Live Time}. During the flight, the total time and the live time $T$ 
(i.e., the time during which the data acquisition was available for triggers)
were measured
by means of a pair of  48-bit counters incorporated in the housekeeping system onboard,
and providing  better than 4 microseconds resolution.
The selected data set amounts to a live time of 1,454,802 seconds,
close to 75\% of the total time of data taking.

{\em Reconstruction efficiency}. The overall efficiency $\epsilon$, including the 
trajectory reconstruction and the charge selection efficiencies,
is estimated from MC simulations as a function 
of the particle energy.
This efficiency curve is characterized by a steep rise below 2 TeV 
and reaches a constant value at energies $>$3 TeV for all nuclei. 
%%%%%
The rise is caused by the high sparsification threshold ($\sim$10 MeV)
of the CAL cells, which limits the CAL tracking accuracy at low energy
and consequently 
the charge identification capability of the instrument.
%%%%
For C, O, Ne, Mg, and Si,
the plateau value in the efficiency curve is about 80\% when
considering only the ``golden'' events, 
and decreases to 65\% when including also 
events at large angle within the SCD-CAL acceptance. 
Instead, the more conservative charge cuts applied to select N and Fe samples
(see Section \ref{chargepar}) yield a reduced plateau efficiency
of about 65\% and 35\% (for ``golden'' events), respectively.
%%%%%%%%%%%%%%%%%%%%%%%%%%%%%%%%%%%%%%%%%%%%%%%%%%%%%%%%%%%%%%%%%%%%%%%
\section{Results}
\begin{figure}
\epsscale{1.0}
\plotone{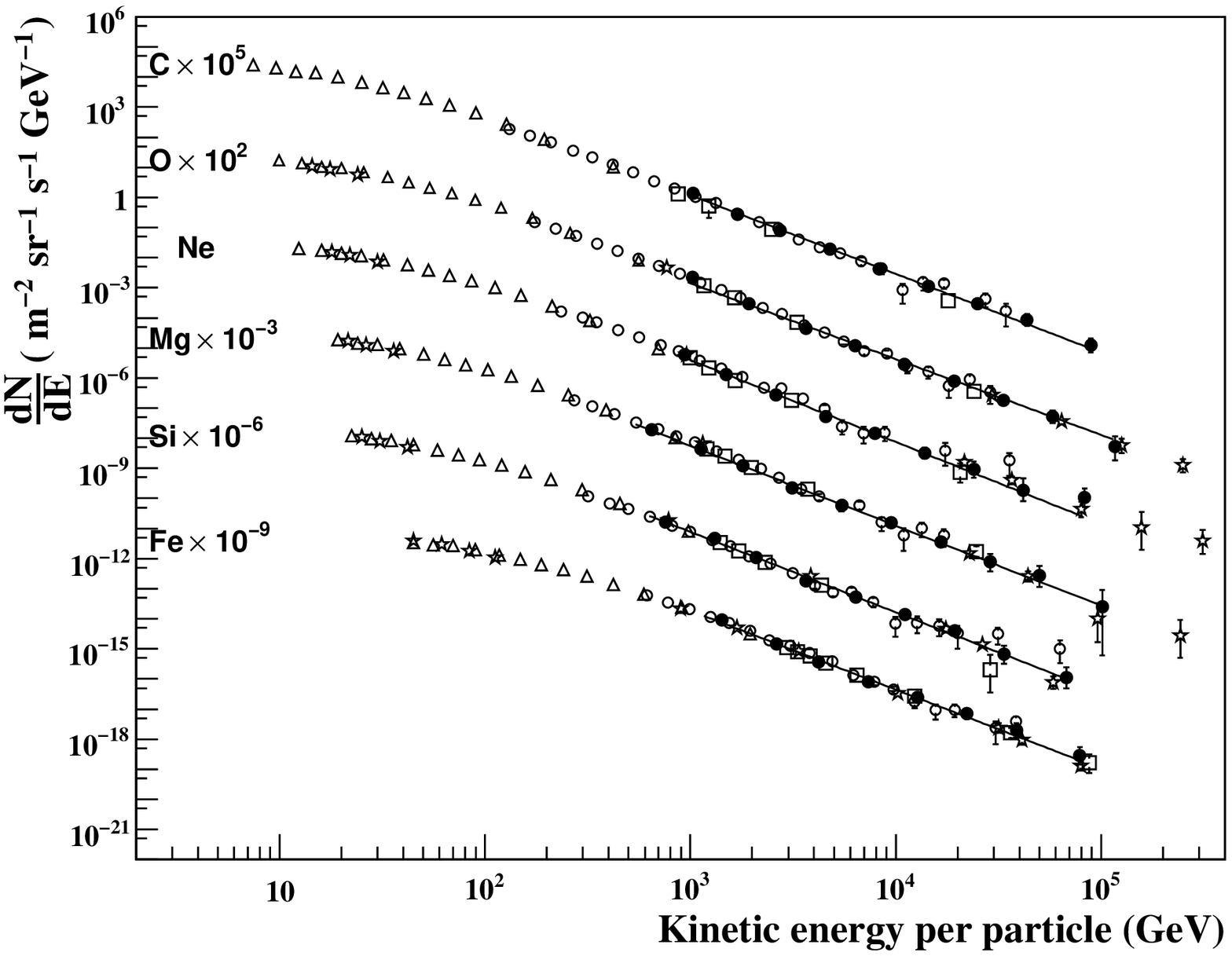}
\caption{Differential intensity as a function of the kinetic energy per particle
for cosmic-ray nuclei C, O, Ne, Mg, Si and Fe, respectively.
CREAM-II data points (filled circles)
are compared with previous observations from:
{\em HEAO 3} C2 \citep{HEAO}, triangles; CRN \citep{CRN}, squares; ATIC-2 \citep{ATIC}, open circles;
TRACER \citep{TRACER}, stars. 
The solid line represents a power-law fit to the CREAM-II data. 
The fitted values of the spectral indices  are listed in Table \ref{tb2}. }
\label{spectra1}
\end{figure}

\begin{figure}
\epsscale{.85}
\plotone{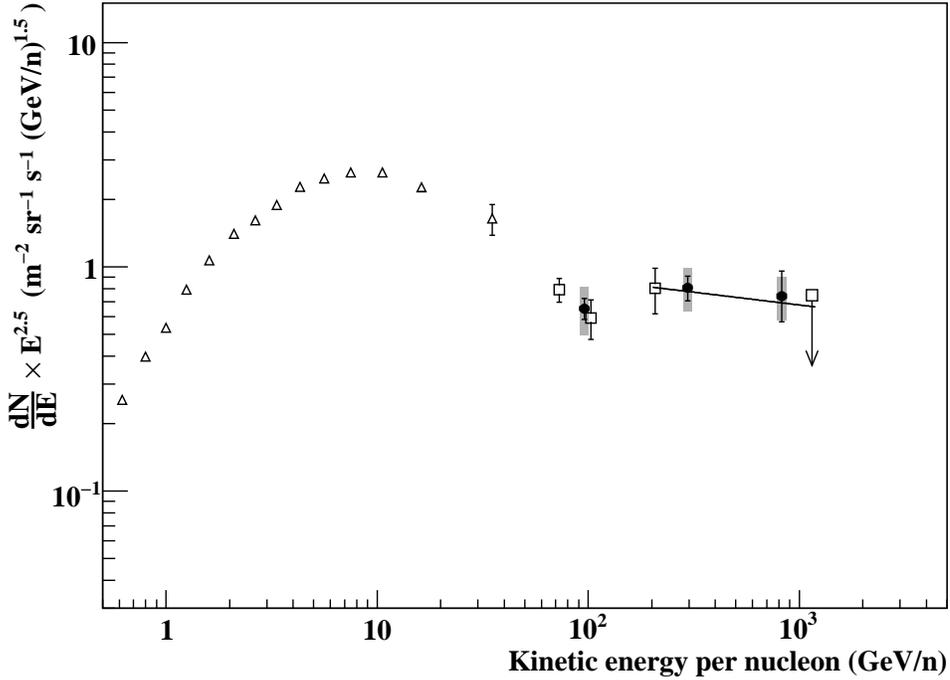}
\caption{ Measurements of the nitrogen energy spectrum (multiplied by E$^{2.5}$).  
CREAM-II results (filled circles) 
are compared with previous observations by
{\em HEAO 3} C2 \citep{HEAO}  (triangles) and CRN \citep{CRN2}  (squares). 
The error bars in CREAM-II data represent only the statistical errors,
while the gray bands show the systematic uncertainties in the differential intensity.
The solid line represents a power-law fit to the combined data at energies $>$ 200 GeV/$n$.
The fitted spectral index is $2.61\pm0.10$ ($\chi^2$/ndf = 0.4/2).
}
\label{Nspectrum}
\end{figure}

The differential intensities at the TOA
as measured by CREAM-II for the elements C, N, O, Ne, Mg, Si, and Fe 
are given  as a function of the kinetic energy  per nucleon in Table \ref{tb1}, 
together with the corresponding  number of raw observed particles.
The quoted error bars  
are  due to the counting statistics only: the upper and lower limits
are computed as  84\% confidence limits for Poisson distributions as described by \citet{gehrels}.\\
The energy spectra of the major primary CR nuclei from carbon to iron  are plotted as a function of the 
kinetic energy per particle in Figure \ref{spectra1}. 
To emphasize the spectral differences, the differential
intensities are multiplied by $\hat{E}^{2.5}$ and plotted 
as a function of the kinetic energy per nucleon in Figure~\ref{spectra2}.
The  error bars shown in the figures represent the sum in quadrature  of the statistical and
systematic uncertainties. The assessment of systematics is discussed in detail in the following section.
The particle energy range covered by CREAM-II 
extends from around 800 GeV up to 100 TeV.
The absolute intensities are presented without any arbitrary normalization to previous data
and cover a range of six decades.\\
The energy spectrum of nitrogen is shown in Figure~\ref{Nspectrum}. 
The statistical and systematic
errors in the differential intensity 
are shown separately to point out that  
the systematic uncertainties 
from the corrections for secondary particle production
in the atmosphere and the instrument,
are particularly relevant for this measurement. In practice, they 
impose a limit on the maximum energy at which measurements of  nitrogen 
in CRs might be pursued on a balloon flight.\\
The elemental abundances with respect to oxygen are calculated from
the differential intensities at the TOA and
are shown as a function of energy in Figure~\ref{ratio}. 
Energy bins at high energy  are merged together 
in order to reduce the statistical error in the ratio to an acceptable level.
%%%%%%%%%%%%%%%%%%%%%%%%%%%%%%%%%%%%%%%%%%%%%%%%%%%%%%%%%%%%%%%%
\subsection{Estimate of the systematic errors}\label{syspar}
The main systematic uncertainties in the differential intensity stem from 
the reconstruction algorithm as well as
 from the TOI and TOA corrections.
To estimate the first, 
we scan
a range of thresholds around the reference value of each selection cut and derive the corresponding 
fluxes. 
In particular, 
the analysis procedure has been repeated 
by varying the fiducial area of the SCD and CAL, 
the coherence level between the signals of the SCD pixels
crossed by the CR particle, and the limits of  charge selection cut for each element. 
Comparing the results with the reference, 
a systematic fractional error in the differential intensity, arising  from the reconstruction 
procedure, is estimated to be of order 10\% below a particle energy of 3 TeV and 5\% above.\\
%%%%%%%%%%%%%%%%%%%%%%
The second source of systematic errors comes from the uncertainties in 
the nucleus-nucleus charge-changing cross sections 
used to calculate
the instrument and atmospheric corrections.
While there is a general agreement between 
different parameterizations of the total cross sections, leading to very similar
values of the survival fractions of CRs in the atmosphere and in the instrument
(Figures \ref{TOI} and \ref{TOA}), the partial charge-changing cross sections 
reported in the literature are measured with quoted errors of order 10\%-15\%, 
depending on the specific nucleus-nucleus interaction, and 
can differ by up to $\sim\, $30\%. This may result in significantly different 
values of the off-diagonal elements 
$F_{Z_2 \rightarrow Z_1}$ in the transport matrix.
For instance, the values we calculated for the 
fraction of oxygen nuclei which spallate in the atmosphere producing 
nitrogen nuclei ($F_{O \rightarrow N}$), are respectively: 1.7\% from the partial cross-section model of \citet{tsao}; 
1.9\%  according to the parameterization of \citet{nilsen}; 1.5\%  from FLUKA \citep{fluka}.
In order to estimate how these uncertainties in the corrections for secondary nuclei production
in the atmosphere and  the instrument
affect our final results,
we developed the following method. 
Using the same
sample of analyzed data for each element, 
we varied the values of the charge-changing probabilities in the range allowed by the different models
and derived, for each set of values, the corresponding energy spectrum.
By comparing the results with the reference spectrum, 
the systematic error for the uncertainty in correcting 
the differential intensity to the top of instrument is estimated to be 2\%
for the primary elements. For nitrogen, a 15\% error is assigned because of the large contamination 
of O nuclei which spallate into N. 
Similar values  are estimated for the systematic error 
from the uncertainties in the atmospheric
secondary corrections.\\
%%%%%%%%%%%%%%%%%%
The systematic error in the energy scale, as measured by the CAL,   
derives from the accuracy with which the simulated CAL response 
corresponds to the real behavior of the calorimeter at energies not covered
by beam test measurements.
We estimate this accuracy at a 5\% level, which corresponds, given the CAL linearity in the 
energy range considered in this work, to an energy-scale uncertainty of 5\%.
%%%%%%%%%%%%%%%%%%%%%%%%%%%%%%%%%%%%%%%%%%%%%%%%%%%%%%%%%%%%%%%%%%%%%%%
\section{Discussion} 
\begin{figure}
\epsscale{1}
\plotone{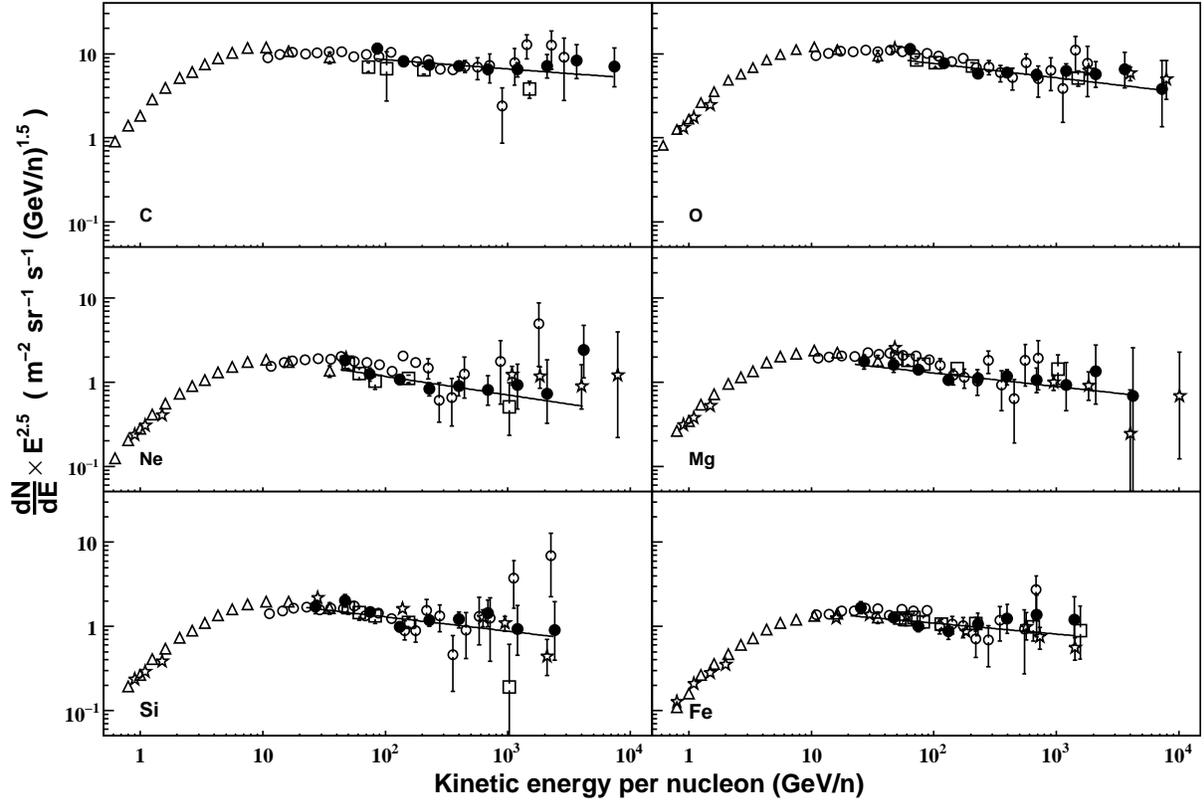}
\caption{Energy spectra  (per nucleon) for the elements  C, O, Ne, Mg, Si and Fe, respectively.
The differential intensities are multiplied by E$^{2.5}$.
CREAM-II results (filled circles) 
are compared with previous observations by:
{\em HEAO 3} C2 \citep{HEAO}, triangles; CRN \citep{CRN}, squares; ATIC-2 \citep{ATIC}, open circles; TRACER \citep{TRACER}, stars.
The solid line represents a power-law fit to the CREAM-II data. } 
\label{spectra2}
\end{figure}

\begin{figure}
\epsscale{1}
\plotone{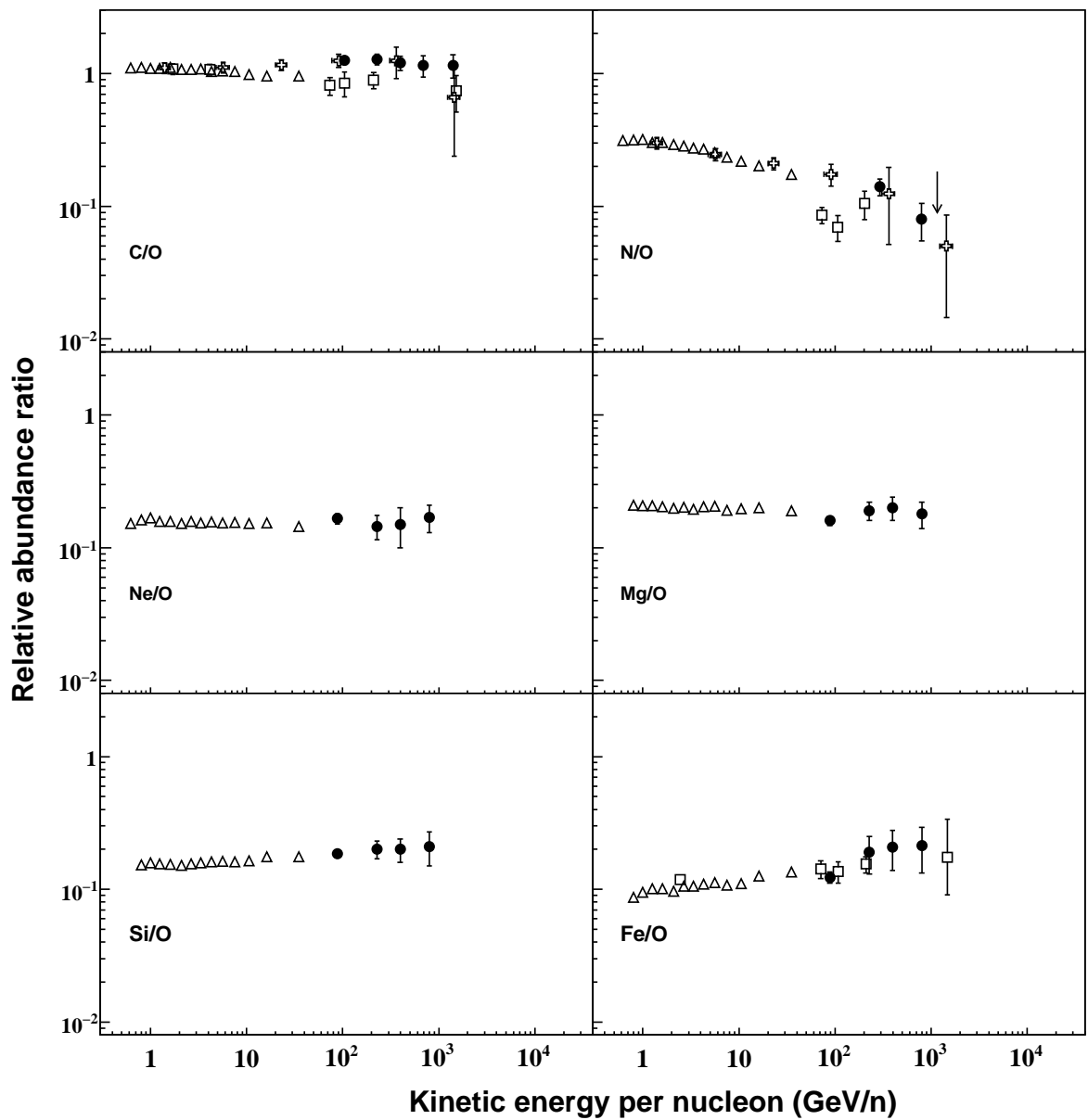}
\caption{Measurements of the relative abundance ratios of C, N, Ne, Mg, Si and Fe, respectively,
 to O as a function of energy. 
CREAM data from the first (open crosses) and second (filled circles) flight are compared with 
previous observations by {\em HEAO 3} C2 \citep{HEAO}  (triangles) and CRN \citep{CRN2}  (squares). 
The arrow in the N/O plot is an upper limit set by CRN at 1.1 TeV/$n$.}
\label{ratio}
\end{figure}

\begin{figure}
\epsscale{.9}
\plotone{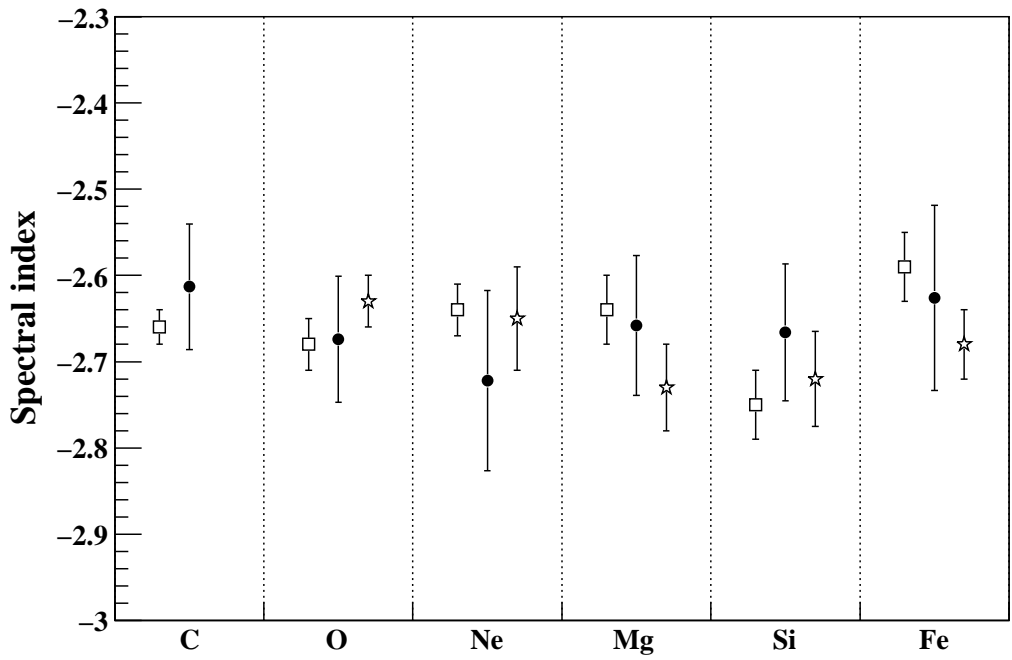}
\caption{
The fitted spectral indices (Table~\ref{tb2}) from CREAM-II data (filled circles)
are compared 
with the spectral indices of a power-law fit to: (stars) 
the combined CRN and TRACER data above 20 GeV/$n$ \citep{TRACER};
(open squares) a compilation of direct and indirect measurements 
 \citep{Wiebel}.}
\label{spectralindex}
\end{figure}

The CREAM-II results for primary nuclei
are found   to be in good agreement with previous measurements of
 space-based  ({\em HEAO 3} C2 \citep{HEAO}, CRN \citep{CRN}) and 
balloon-borne (ATIC-2 \citep{ATIC}, TRACER \citep{TRACER}) 
experiments (see Figures~\ref{spectra1} and \ref{spectra2}).
All the elements appear to have the same spectral shape,
which can be described by a single power law in energy $E^{-\gamma}$.
The spectral indices fitted to our data 
(Table \ref{tb2}) are very similar, 
indicating
that the intensities of the more abundant (evenly) charged heavy elements
have nearly the same energy dependence. 
Our observations, based on a calorimetric measurement of the CR energy, 
confirm the results recently reported by the TRACER collaboration (Figure~\ref{spectralindex}), 
using  completely different techniques (Cherenkov, specific ionization in gases and 
transition radiation) to determine  the particle energy.
The weighted average of our fitted spectral indices is
 $\bar{\gamma} = 2.66 \pm 0.04$, consistent, within  error,
with the value of $2.65 \pm 0.05$
obtained from a fit to the combined CRN and TRACER data \citep{TRACER}.
The great similarity of the spectral indices suggests that 
the same mechanism is responsible for the source acceleration  of these primary heavy nuclei.\\
Although all the elemental spectra can be fitted to a single power law in energy, 
hardening (flattening) of the spectra above $\sim\,$200 GeV/$n$ is apparent 
from the overall trend of the data (Figure~\ref{spectra2}). 
A detailed analysis to investigate possible features in the spectra is in progress 
to evaluate the significance of a change in the spectral index. \\
CREAM-II data extend the energy range spanned by the previous elemental abundance measurements
 up to $\sim $ 1 TeV/n (Figure~\ref{ratio}). 
The C/O ratio measured by CREAM-II is consistent with the one reported by CREAM-I \citep{Ahn2008} 
within the overlapping energy region covered by the two flights and slightly higher than the CRN result.
The Ne/O and Mg/O ratios confirm, up to 800 GeV/$n$, 
the nearly flat trend of {\em HEAO 3} C2 low-energy data, while
the fractions of Si and Fe with respect to O seem to increase with energy. 
This could represent the clue of
a possible change in composition 
of CRs at high energies with an
enhancement of heavier elements, as expected from SNR shock diffusive acceleration theories.
Nevertheless, the statistics of the data are  too limited for a definitive conclusion at this time.

%%%%%%%%%%%%%
Unlike the primary heavy nuclei, nitrogen is 
mostly produced by spallation in the interstellar medium,
but it also has  a  primary contribution of order 10\%, as recently measured by CREAM-I \citep{Ahn2008}. 
The measurement of  nitrogen intensity at high energy is challenging
because it requires an excellent charge separation between
N nuclei and the much more abundant C and O neighbors, as well as 
a complete understanding of the corrections for secondary N particle 
production in the atmosphere and the instrument.
The nitrogen data collected by CREAM-II are more statistically 
significant at high energy than any previous observation.
We notice that, above 100 GeV/$n$, the N spectrum
flattens out from the steep decline which characterizes the energy range 10-100 GeV/$n$  (Figure~\ref{Nspectrum}).
The combined CRN and CREAM-II data at energies higher than 200 GeV/$n$,
are well fitted to a power-law in energy with spectral index $2.61 \pm 0.10$, 
consistent within the error with the average value $\bar{\gamma}$ of the primary nuclei.
This supports the hypothesis of the presence of two components in the cosmic nitrogen flux.
In fact, since the escape pathlength decreases rapidly with energy, 
the secondary nitrogen component is expected to become negligible  
at high energy, where only the primary component should survive. This might result in a change of the 
spectral slope, as observed.\\
As a cross-check to the  measurement reported here, we calculate the
relative abundance ratio of N/O 
and compare it with the CREAM-I observation (Figure~\ref{ratio}), 
in which the energy of CRs was measured with a
transition radiation detector. 
A N/O ratio 
 (at the TOA) equal to  $0.080 \pm 0.025\, $(stat.)$\, \pm 0.025\,$(sys.)
is measured
at an energy of $\sim\,$800 GeV/$n$, in good agreement with recent results from
CREAM-I \citep{Ahn2008}. % and a previous analysis with CREAM-II data \citep{AhnSubmit}.
%%%%%%%%%%%%%%%%%%%%%%%%%%%%%%%%%%%%%%%%%%%%%%%%
\section{Conclusions} 
CREAM-II  carried out measurements of high-$Z$ CR nuclei
with an excellent charge resolution and  a reliable energy determination.
The use of a dual layer of silicon sensors allowed us to unambiguously identify each individual element
up to iron and to reduce the background of misidentified nuclei.
We demonstrated the feasibility of measuring 
 the particle energy up to hundreds of TeV by using  a thin ionization sampling calorimeter,
with sufficient resolution to reconstruct accurately the CR spectra. \\
The energy spectra of the major primary heavy nuclei from C to Fe
were measured up to $\sim\,$10$^{14}$ eV and found to agree well with earlier direct measurements.
All the spectra follow a power law in energy with a remarkably similar  
 spectral index $\bar{\gamma} = -2.66 \pm 0.04$, as  is plausible if they have
the same origin and  share the same acceleration and propagation processes. \\
A new measurement of the nitrogen intensity in an energy region so far experimentally unexplored 
indicates a harder  spectrum than at lower energies, 
supporting the idea that nitrogen has  secondary as well as primary contributions.
These results 
provide new clues for understanding the CR acceleration and propagation mechanisms,  
%open a new window in the exploration of the TeV/n region of charged cosmic rays, but 
but further accurate observations with high statistics beyond 10$^{14}$ eV are needed to finally unravel 
the mystery of CR origin.
%the CR origin and acceleration mechanism. 

\acknowledgments
This work is supported by NASA research grants to the University of Maryland, 
Penn State University, and Ohio State University, 
by the Creative Research Initiatives program (RCMST) of MEST/NRF in Korea, and INFN and PNRA in Italy. The authors
greatly appreciated the support of NASA/WFF and NASA/GSFC, 
Columbia Scientific Balloon Facility, National Science Foundation's Office
of Polar Programs and Raytheon Polar Services Company for the
successful balloon launch, flight operation and payload recovery in Antarctica.
%%%%%%%%%%%%%%%%%%%%%%%%%%%%%%%%%%%%%%%%%%%%%%%%%%%%%%%%%%%%%%%%%%%%%%%

%%%%%%%%%%%%%%%%%%%%%%%%%%%%%%%%%%%%%%%%%%%%%%%%%%%%%%%%%%%%%%%%%%%%%%%
\clearpage
\begin{deluxetable}{lcccc}
\newcommand{\m}{\hphantom{$8$}}
\renewcommand{\arraystretch}{1.0} % enlarge line spacing
\tablewidth{0pt}
\tablecaption{Differential Intensities Measured with CREAM-II \label{tb1}}
\tablehead{
\colhead{Element}        & \colhead{Energy Range}   &
\colhead{Kinetic Energy}          & \colhead{No. of Events}  &
\colhead{Intensity%\tablenotemark{a}
}   \\ \colhead{}    &
\colhead{(GeV/$n$)}  & \colhead{$\hat{E}$ (GeV/$n$)}  &
\colhead{} & \colhead{(m$^{2}$ sr s GeV/$n$)$^{-1}$} }
\startdata
C         &\m\m66 - \m117 &\m\m86.2      & \m171     &(1.67 $\pm$ 0.13)$\times$10$^{-4}$ \\
$Z$=6       &\m117 - \m175 &\m141.6        & \m139     &(3.37 $\pm$ 0.29)$\times$10$^{-5}$ \\
          &\m175 - \m304 &\m228.2        & \m215     &(9.42 $\pm$ 0.64)$\times$10$^{-6}$ \\
          &\m304 - \m529 &\m397.2        & \m134     &(2.28 $\pm$ 0.20)$\times$10$^{-6}$ \\
          &\m529 - \m919 &\m691.0        & \m\m66    &(5.17 $\pm$ 0.63)$\times$10$^{-7}$ \\
          &\m919 - 1597 &1201.5          & \m\m29    &(1.31 $\pm$ 0.24)$\times$10$^{-7}$ \\
          &1597 - 2776 &2088.7           & \m\m13    &(3.6 $^{+1.3}_{-1.0}$)$\times$10$^{-8}$ \\
          &2776 - 4824 &3630.5           & \m\m\m7   &(1.04 $^{+0.58}_{-0.40}$)$\times$10$^{-8}$ \\
          &4284 -\m \m\m\m\m &7415.1     & \m\m\m5   &(1.49 $^{+0.98}_{-0.63}$)$\times$10$^{-9}$ \\
\hline
\dataset
N         &\m\m56 - \m175   &\m\m95.7    & \m\m86      &(7.28 $\pm$ 0.79)$\times$10$^{-6}$ \\
$Z$=7       &\m175 - \m529    &\m295.3     & \m\m63      &(5.39 $\pm$ 0.68)$\times$10$^{-7}$ \\
          &\m529 - \m\m\m\m &\m826.3     & \m\m19      &(3.8 $^{+1.1}_{-0.9}$)$\times$10$^{-8}$ \\
\hline
\dataset
O         &\m\m49 - \m\m87 &\m\m64.0     & \m224     &(3.48 $\pm$ 0.23)$\times$10$^{-4}$ \\
$Z$=8       &\m\m87 - \m175 &\m120.8       & \m371     &(4.82 $\pm$ 0.25)$\times$10$^{-5}$ \\
          &\m175 - \m304 &\m228.0        & \m241     &(7.33 $\pm$ 0.47)$\times$10$^{-6}$ \\
          &\m304 - \m529 &\m396.9        & \m127     &(1.90 $\pm$ 0.17)$\times$10$^{-6}$ \\
          &\m529 - \m919 &\m690.5        & \m\m57    &(4.51 $\pm$ 0.60)$\times$10$^{-7}$ \\
          &\m919 - 1597 &1200.5          & \m\m25    &(1.25 $\pm$ 0.25)$\times$10$^{-7}$ \\
          &1597 - 2776 &2087.0           & \m\m10    &(2.9 $^{+1.2}_{-0.9}$)$\times$10$^{-8}$ \\
          &2776 - 4824 &3627.4           & \m\m\m6   &(8.3 $^{+4.9}_{-3.3}$)$\times$10$^{-9}$ \\
          &4284 -\m \m\m\m\m &7287.1     & \m\m\m3   &\m(8.4 $^{+9.7}_{-5.4}$)$\times$10$^{-10}$ \\
\hline
\dataset
Ne        &\m\m40 - \m\m58 &\m\m47.0     & \m\m37    &(1.20 $\pm$ 0.20)$\times$10$^{-4}$ \\
$Z$=10      &\m\m58 - \m101 &\m74.8        & \m\m65    &(2.59 $\pm$ 0.32)$\times$10$^{-5}$ \\
          &\m101 - \m175 &\m130.7        & \m\m67    &(5.53 $\pm$ 0.68)$\times$10$^{-6}$ \\
          &\m175 - \m304 &\m227.9        & \m\m36    &(1.06 $\pm$ 0.18)$\times$10$^{-6}$ \\
          &\m304 - \m529 &\m396.7        & \m\m19    &(2.86 $^{+0.84}_{-0.66}$)$\times$10$^{-7}$ \\
          &\m529 - \m919 &\m690.0        & \m\m\m8   &(6.5 $^{+3.1}_{-2.2}$)$\times$10$^{-8}$ \\
          &\m919 - 1597 &1199.8          & \m\m\m4   &(1.8 $^{+1.4}_{-0.9}$)$\times$10$^{-8}$ \\
          &1597 - 2776&2085.6            & \m\m\m2   &(3.7 $^{+5.6}_{-2.0}$)$\times$10$^{-9}$ \\
          &2776 - \m\m\m\m &4150.9       & \m\m\m3   &(2.2 $^{+2.1}_{-1.2}$)$\times$10$^{-9}$ \\
\hline
\dataset
Mg        &\m\m20 - \m\m40 &\m\m27.0     & \m\m48    &(4.68 $\pm$ 0.67)$\times$10$^{-4}$ \\
$Z$=12      &\m\m40 - \m\m58 &\m\m47.0     & \m\m45    &(1.07 $\pm$ 0.16)$\times$10$^{-4}$ \\
          &\m\m58 - \m101 &\m74.9        & \m103     &(2.91 $\pm$ 0.29)$\times$10$^{-5}$ \\
          &\m101 - \m175 &\m130.8        & \m\m94    &(5.41 $\pm$ 0.56)$\times$10$^{-6}$ \\
          &\m175 - \m304 &\m228.0        & \m\m51    &(1.42 $\pm$ 0.20)$\times$10$^{-6}$ \\
          &\m304 - \m529 &\m397.0        & \m\m26    &(3.72 $\pm$ 0.73)$\times$10$^{-7}$ \\
          &\m529 - \m919 &\m690.6        & \m\m12    &(8.5 $^{+3.3}_{-2.4}$)$\times$10$^{-8}$ \\
          &\m919 - 1597 &1200.8          & \m\m\m4   &(1.9 $^{+1.6}_{-1.0}$)$\times$10$^{-8}$ \\
          &1597 - 2776 &2087.5           & \m\m\m3   &(6.7 $^{+7.1}_{-4.0}$)$\times$10$^{-9}$ \\
          &2776 - \m\m\m\m &4215.9       & \m\m\m1   &\m(6.0 $^{+16.0}_{-5.8}$)$\times$10$^{-10}$ \\
\hline
\dataset
Si        &\m\m20 - \m\m40 &\m\m27.0     & \m\m66     &(4.54 $\pm$ 0.56)$\times$10$^{-4}$ \\
$Z$=14      &\m\m40 - \m\m58 &\m\m47.0     & \m\m64     &(1.35 $\pm$ 0.17)$\times$10$^{-4}$ \\
          &\m\m58 - \m101 &\m74.9        & \m116      &(3.03 $\pm$ 0.28)$\times$10$^{-5}$ \\
          &\m101 - \m175 &\m130.8        & \m\m89     &(5.04 $\pm$ 0.53)$\times$10$^{-6}$ \\
          &\m175 - \m304 &\m228.0        & \m\m54     &(1.50 $\pm$ 0.20)$\times$10$^{-6}$ \\
          &\m304 - \m529 &\m397.0        & \m\m24     &(3.88 $\pm$ 0.79)$\times$10$^{-7}$ \\
          &\m529 - \m919 &\m690.6        & \m\m10     &(1.15 $^{+0.48}_{-0.35}$)$\times$10$^{-7}$ \\
          &\m919 - 1597 &1200.7          & \m\m3      &(1.9 $^{+1.7}_{-0.9}$)$\times$10$^{-8}$ \\
          &1597 - \m\m\m\m &2418.3       & \m\m2      &(3.2 $^{+3.6}_{-1.8}$)$\times$10$^{-9}$ \\
\hline
\dataset
Fe        &\m\m18 - \m\m40 &\m\m25.4     & \m108      &(5.19 $\pm$ 0.56)$\times$10$^{-4}$ \\
$Z$=26      &\m\m40 - \m\m58 &\m\m47.0     & \m\m88     &(8.38 $\pm$ 0.89)$\times$10$^{-5}$ \\
          &\m\m58 - \m101 &\m74.9        & \m\m70     &(2.06 $\pm$ 0.25)$\times$10$^{-5}$ \\
          &\m101 - \m175 &\m131.0        & \m\m27     &(4.46 $\pm$ 0.85)$\times$10$^{-6}$ \\
          &\m175 - \m304 &\m228.2        & \m\m15     &(1.36 $^{+0.46}_{-0.36}$)$\times$10$^{-6}$ \\
          &\m304 - \m529 &\m397.2        & \m\m\m9    &(4.0 $^{+1.8}_{-1.3}$)$\times$10$^{-7}$ \\
          &\m529 - \m919 &\m690.9        & \m\m\m4    &(1.09 $^{+0.79}_{-0.48}$)$\times$10$^{-7}$ \\
         &\m919 - \m\m\m\m &1406.9      & \m\m\m4    &(1.6 $^{+1.4}_{-0.8}$)$\times$10$^{-8}$ \\
\enddata
\end{deluxetable}
%%%%%%%%%%%%%%%%%%%%%%%%%%%%%%%%%%%%%%%%%%%%%%%%%%%%%%%%%%%%%%%%%%%%%%%%%%%%%
\clearpage
\begin{table}
\newcommand{\m}{\hphantom{$8$}}
\renewcommand{\tabcolsep}{1.5pc} 
\renewcommand{\arraystretch}{1.0} 
\begin{center}
\caption{Best fit values of the spectral index $\gamma$  
of each element, as measured by CREAM-II, and $\chi^2$/ndf of the fit\label{tb2}}
\begin{tabular}{ccc}
\tableline\tableline
Element & $\gamma$ & $\chi^2$/ndf \\
\tableline
C\m  & $2.61\pm0.07$ & $3.3/7$\\
O\m  & $2.67\pm0.07$ & $4.8/7$\\
Ne & $2.72\pm0.10$ & 4.2/7\\  
Mg & $2.66\pm0.08$ & 1.9/8\\
Si\m & $2.67\pm0.08$ & 4.9/7\\
Fe\m & $2.63\pm0.11$ & 4.2/6\\
\tableline
\end{tabular}
\end{center}
\end{table}
%%%%%%%%%%%%%%%%%%%%%%%%%%%%%%%%%%%%%%%%%%%%%%%%%%%%%%%%%%%%%%%%%%%%%%%%%%%%%
%%%%%%%%%%%%%%%%%%%%%%%%%%%%%%%%%%%%%%%%%%%%%%%%%%%%%%%%%%%%%%%%%%%%%%%%%%%%%
%%%%%%%%%%%%%%%%%%%%%%%%%%%%%%%%%%%%%%%%%%%%%%%%%%%%%%%%%%%%%%%%%%%%%%%%%%%%%
%%%%%%%%%%%%%%%%%%%%%%%%%%%%%%%%%%%%%%%%%%%%%%%%%%%%%%%%%%%%%%%%%%%%%%%%%%%%%
%%%%%%%%%%%%%%%%%%%%%%%%%%%%%%%%%%%%%%%%%%%%%%%%%%%%%%%%%%%%%%%%%%%%%%%%%%%%%
%%%%%%%%%%%%%%%%%%%%%%%%%%%%%%%%%%%%%%%%%%%%%%%%%%%%%%%%%%%%%%%%%%%%%%%%%%%%%
%%%%%%%%%%%%%%%%%%%%%%%%%%%%%%%%%%%%%%%%%%%%%%%%%%%%%%%%%%%%%%%%%%%%%%%%%%%%%
%%%%%%%%%%%%%%%%%%%%%%%%%%%%%%%%%%%%%%%%%%%%%%%%%%%%%%%%%%%%%%%%%%%%%%%%%%%%%
\clearpage
%%%%%%%%%%%%%%%%%%%%%%%%%%%%%%%%%%%%%%%%%%%%%%%%%%%%%%%%%%%%%%%%%%%%%%%%%%%%%
\clearpage
%%%%%%%%%%%%%%%%%%%%%%%%%%%%%%%%%%%%%%%%%%%%%%%%%%%%%%%%%%%%%%%%%%%%%%%%%%%%%
\clearpage
%%%%%%%%%%%%%%%%%%%%%%%%%%%%%%%%%%%%%%%%%%%%%%%%%%%%%%%%%%%%%%%%%%%%%%%%%%%%%

\begin{thebibliography}{}
\bibitem[Aharonian et al.(2004)]{HESS} Aharonian, F. et al., 2004, Nature, 432, 75  
\bibitem[Ahn et al.(2004)]{Ahn2004} Ahn, H.S. et al., 2004, Proc. of 11$^{\text{th}}$ International Conference on Calorimetry
in Particle Physics, (Perugia, Italy), 2004, 532
\bibitem[Ahn et al.(2006)]{Ahn} Ahn, H.S. et al., 2006, Nucl. Phys. B (Proc. Suppl.), 150, 272
\bibitem[Ahn et al.(2008)]{Ahn2008} Ahn, H.S. et al., 2008, Astropart. Phys. 30, 133 
\bibitem[Ahn et al.(2009)]{Ahn2009} Ahn, H.S. et al., 2009, Nucl. Instrum. Methods A, 602, 525
%\bibitem[Ahn et al.(submitted)]{AhnSubmit} Ahn, H.S. et al., submitted to \apj, ``Measurements of the relative abundances of high-energy cosmic-ray nuclei in the TeV/nucleon region''
\bibitem[Ave et al.(2008)]{TRACER} Ave, M. et al., 2008, \apj, 678, 262 
\bibitem[Bell(1978)]{Bell} Bell, A.R. 1978, MNRAS, 182, 443 
\bibitem[Berezhko(1996)]{Berezhko} Berezhko, E.G. 1996, Astropart. Phys. 5, 367 
\bibitem[Binns et al.(1988)]{binns} Binns, W.R. et al., 1988, \apj, 324, 1106
\bibitem[Engelman et al.(1990)]{HEAO} Engelmann, J.J. et al., 1990, A\&A, 233, 96
\bibitem[Fass\`{o} et al.(2005)]{fluka} Fass\`{o}, A., Ferrari, A., Ranft, J., \& Sala, P.R.,
``FLUKA: A Multi-Particle Transport Code'', CERN-2005-10 (2005), INFN/TC\_05/11, SLAC-R-773
\bibitem[Gehrels(1986)]{gehrels} Gehrels, N. 1986, \apj, 303, 336
\bibitem[Hillas(2005)]{hillas} Hillas, A.M. 2005, J. Phys. G: Nucl. Part. Phys., 31, R95
\bibitem[H\"{o}randel(2003)]{horandel} H\"{o}randel, J.R. 2003, Astropart. Phys., 19, 193
\bibitem[H\"{o}randel(2004)]{horandel2} H\"{o}randel, J.R. 2004, Astropart. Phys., 21, 241
\bibitem[Kox et al.(1987)]{kox} Kox, S. et al., 1987, Phys. Rev. C, 35, 1678
\bibitem[Lafferty \& Wyatt(1995)]{LW} Lafferty, G.D., \& Wyatt, T.T. 1995, Nucl. Instrum. Methods A, 355, 541
\bibitem[Lagage \& Cesarsky(1983)]{LC} Lagage, P.O., \& Cesarsky, C.J. 1983, A\&A, 125, 249
\bibitem[Longair(1994)]{longair}  Longair, M.S. 1994, High Energy Astrophysics 2, Cambridge Univ. Press
\bibitem[Marrocchesi et al.(2005)]{Marrocchesi2} Marrocchesi, P.S. et al., 2004, Nucl. Instrum. Methods A, 535, 143
\bibitem[Marrocchesi et al.(2008)]{Marrocchesi} Marrocchesi, P.S. et al., 2008, Adv. Space Res., 41, 2002
\bibitem[Nam et al.(2007)]{Nam} Nam, S. et al., 2007, IEEE Trans. Nucl. Sci., 54, 1743
\bibitem[Nilsen et al.(1995)]{nilsen} Nilsen, B.S. et al., 1995, Phys. Rev. C, 52, 3277
\bibitem[M\"{u}ller et al.(1991)]{CRN} M\"{u}ller, D. et al., 1991, \apj, 374, 356
\bibitem[Panov et al.(2006)]{ATIC} Panov, A.D. et al., 2006, Adv. Space Res., 37, 1944
\bibitem[Parizot et al.(2004)]{parizot} Parizot, E., Marcowith, A., van der Swaluw, E., Bykov, A.M., \& Tatischeff, V., 2004, A\&A, 424, 747
\bibitem[Park et al.(2007)]{Park} Park, I.H. et al., 2007, Nucl. Instrum. Methods A, 570, 286
\bibitem[Seo et al.(2004)]{seo} Seo, E.S. et al., 2004, Adv. Space Res., 33, 1777
\bibitem[Sivher et al.(1993)]{sivher} Sivher, L. et al., 1993, Phys. Rev. C, 47(3), 1225
\bibitem[Swordy et al.(1990)]{CRN2} Swordy, S.P. et al., 1990, \apj, 349, 625
\bibitem[Tsao et al.(1993)]{tsao} Tsao, C.H. et al., 1993, Phys. Rev. C, 47(3), 1257
\bibitem[Yanasak et al.(2001)]{yanasak} Yanasak, N.E. et al., 2001, \apj, 563, 768
\bibitem[Zei et al.(2008)]{Zei} Zei, R. et al., 2008, in Proc. of 30$^{\text{th}}$ International Cosmic Ray Conference (ICRC), Vol. 2, 23 
\bibitem[Wiebel-Sooth et al.(1998)]{Wiebel} Wiebel-Sooth, B. et al., 1998, A\&A, 330, 389
\end{thebibliography}
\end{document}